\pdfoutput=1
\PassOptionsToPackage{table}{xcolor}
\documentclass[12pt]{article}

\usepackage{microtype}
\usepackage[utf8]{inputenc}
\usepackage[T1]{fontenc}

\usepackage{amsmath,amssymb}

\usepackage{graphicx}

\usepackage{listings}
\usepackage{algorithm}
\usepackage{algorithmic}

\usepackage{hyperref}
\usepackage{enumitem}
\usepackage{xcolor}

\usepackage{tabularx}
\usepackage{booktabs}    
\usepackage{multirow}    
\usepackage{makecell}    
\usepackage{longtable}   
\usepackage{pdflscape}
\usepackage{subcaption}
\usepackage{array}

\usepackage{tcolorbox}
\tcbuselibrary{listings, breakable, skins}

\usepackage{natbib}

\usepackage[margin=0.8in]{geometry}

\newtcolorbox{PromptBox}[1][]{
	breakable,
	enhanced,
	colback=gray!5,
	colframe=black!40,
	title=#1,
	fonttitle=\bfseries,
	boxrule=0.4pt,
	arc=2pt,
	top=2pt,
	bottom=2pt,
	left=4pt,
	right=4pt,
	width=\textwidth
}

\expandafter\def\expandafter\normalsize\expandafter{%
	\normalsize%
	\setlength\abovedisplayskip{4pt}%
	\setlength\belowdisplayskip{4pt}%
	\setlength\abovedisplayshortskip{0pt}%
	\setlength\belowdisplayshortskip{2pt}%
}

\title{The Vulnerability of LLM Rankers to Prompt Injection Attacks}

\author{
	Yu Yin$^{1}$ \quad
	Shuai Wang$^{1}$ \quad
	Bevan Koopman$^{1,2}$ \quad
	Guido Zuccon$^{1}$ \\
	\normalsize $^1$The University of Queensland, Brisbane, Australia \\
	\normalsize $^2$CSIRO, Brisbane, Australia \\
	\normalsize \texttt{y.yin1@uq.edu.au, shuai.wang2@uq.edu.au} \\
	\normalsize \texttt{bevan.koopman@csiro.au, g.zuccon@uq.edu.au}
}

\date{}

\begin{document}
	\maketitle
	
	\begin{abstract}
	Large Language Models (LLMs) have emerged as powerful re-rankers. Recent research has however showed that simple prompt injections embedded within a candidate document (i.e., jailbreak prompt attacks) can significantly alter an LLM's ranking decisions. 
 While this poses serious security risks to LLM-based ranking pipelines, the extent to which this vulnerability persists across diverse LLM families, architectures, and settings remains largely under-explored.
In this paper, we present a comprehensive empirical study of jailbreak prompt attacks against LLM rankers. 
We focus our evaluation on two complementary tasks: (1) \textit{Preference Vulnerability Assessment}, measuring intrinsic susceptibility via attack success rate (ASR); and (2) \textit{Ranking Vulnerability Assessment}, quantifying the operational impact on the ranking's quality (nDCG@10). We systematically examine three prevalent ranking paradigms (pairwise, listwise, setwise) under two injection variants: decision objective hijacking and decision criteria hijacking. Beyond reproducing prior findings, we expand the analysis to cover vulnerability scaling across model families, position sensitivity, backbone architectures, and cross-domain robustness. Our results characterize the boundary conditions of these vulnerabilities, revealing critical insights such as that encoder-decoder architectures exhibit strong inherent resilience to jailbreak attacks. We publicly release our code and additional experimental results at \url{https://github.com/ielab/LLM-Ranker-Attack}.
	\end{abstract}

	\section{Introduction}
	Large language models (LLMs) have emerged as powerful components in information retrieval (IR) systems ~\cite{hou2024large, lin2025can, li2024survey, wu2024survey,zuccon2025r2llms}. They often serve as effective re-rankers, achieving performance that frequently surpasses traditional ranking methods in assessing query-document relevance ~\cite{zhuang2024setwise, sun2023chatgpt, zhang2024two,zuccon2025r2llms}. 
Despite their impressive capabilities, a concerning vulnerability has surfaced: LLM-based rankers can be compromised through simple adversarial prompt injection attacks. 
Specifically, \citet{qian2025ranking} demonstrated that jailbreak prompts, developed focusing on task substitution and ranking criterion manipulation, can manipulate LLM rankers.

\citet{qian2025ranking} identified a critical weakness which emerges during multi-document comparison tasks. In these scenarios, LLMs must simultaneously evaluate relative relevance across multiple documents while maintaining fidelity to ranking instructions, a context where they exhibit unique susceptibility to manipulation. The vulnerability even extends across multiple state-of-the-art LLMs, including models from the Qwen3 family \cite{yang2025qwen3}, LLaMA-3.3-70B \cite{grattafiori2024llama}, and GPT-4.1-mini \cite{achiam2023gpt}. Moreover, contrary to the expectation that larger model size ensures greater robustness, their findings suggest that larger and more capable LLMs exhibit greater susceptibility. These vulnerabilities pose substantial security risks, potentially allowing adversarial actors to attack rankers quality, disseminate misinformation, and erode trust in LLM-based IR systems \cite{zou2025poisonedrag, nestaas2024adversarial}. This phenomenon further raises serious concerns regarding the trustworthiness and robustness of LLM-based ranking systems.

However, the experimental setup and evaluation metrics in the original study operate under specific conditions that not fully represent real-world scenarios. While \citet{qian2025ranking} conducted two case studies to simulate attacks, several inconsistencies remain. First, in a web search scenario they observed that LLaMA-3-70B was more robust than LLaMA-3-8B, which contradicts their main experimental conclusion that larger models are more vulnerable. Second, their evaluation on the TREC DL datasets~\cite{Craswell2019TrecDl}  was limited to two LLaMA-3 models and did not specify critical details such as the attack method variation or ranking paradigm. 
In addition, the original study conducted limited experiments on potential interesting factors such as attack positioning (front vs. end-of-passage), backbone architectural differences (e.g., encoder-decoder vs. decoder-only), and robustness for domain-specific search.

To address these gaps, this paper aims to conduct a comprehensive empirical investigation into the effectiveness of jailbreak prompt attacks under more practical conditions, while simultaneously analyzing the underlying factors contributing to LLM ranker vulnerabilities. 

Specifically, our investigation is organized into two sequential tasks: (1) \textbf{Preference Vulnerability}, which evaluates the fundamental model susceptibility by focusing on disruption of ranker preferences, for which we reproduce and extend the original experiments across diverse scales, positions, and architectures; and (2) \textbf{Ranking Vulnerability}, which examines how these attacks translate into nDCG@10 degradation, for which we introduce a series of new experiments and findings. Table \ref{tab:research_dimensions} provides a synthesized overview of our research directions and key empirical findings in direct comparison with the original study \cite{qian2025ranking}, categorizing outcomes as confirmed, refined, or newly discovered.

\begin{table*}[t]
\centering
\caption{The research directions explored in this paper, along with a summary of the key findings. \vspace{-10pt}}
\label{tab:research_dimensions}
\footnotesize
\renewcommand{\arraystretch}{1.2}
\begin{tabularx}{\columnwidth}{@{} >{\raggedright\arraybackslash}p{2.5cm} X X @{} }
\toprule
\textbf{Direction} & \textbf{Task1: Preference Vulnerability} & \textbf{Task2: Ranking Vulnerability} \\ \midrule
\textbf{Vulnerability Scaling} & \textbf{Confirmed:} Larger LLMs are generally more vulnerable & \textbf{Refined:} Larger and \textbf{more capable} LLMs are more vulnerable \\ \midrule
\textbf{Position Sensitivity} & \textbf{Confirmed:} ASR gaps exist across positions &  \textbf{Discovered:} Position makes statistically significant difference \\ \midrule
\textbf{Architectural Divergence} & \textbf{Discovered:} Encoder-decoder models show great robustness & \textbf{Discovered:} Encoder-decoder prevents degradation \\ \midrule
\textbf{Domain Robustness} & \textbf{Discovered:} Vulnerability transcends domains& \textbf{Discovered:} High ASR $\neq$ severe nDCG@10 degradation. \\ \bottomrule
\end{tabularx}
\end{table*}

	\section{Related Works}
	This section provides a background on LLM-based ranking~\cite{zuccon2025r2llms} and explores emerging vulnerabilities within LLM-powered retrieval and ranking systems.

\subsection{LLM-based Text Ranking}

LLMs have been extensively adapted for text ranking, primarily through four distinct ranking paradigms ~\cite{sun2025investigation}:

\textbf{Pointwise ranking} \cite{liang2022holistic, sachan2022improving}: The LLM predicts a relevance score for a single query-document pair. While effective, this often requires access to output logits to compute likelihood scores, which restricts its utility for proprietary models where such logit access is unavailable. We do not consider pointwise methods as the attacks by Qian et al. are explicitly targeted at disrupting pair/list/set preferences.

\textbf{Pairwise ranking} \cite{qin2024large}: This approach reduces the ranking task to a binary classification, where the LLM compares two documents for a query to determine the more relevant one. Despite its precision, it suffers from poor scalability due to the large complexity of comparing all document pairs during inference.

\textbf{Listwise ranking} \cite{ma2023zero, pradeep2023rankvicuna, sun2023chatgpt}: The LLM receives a query and with a list of candidate documents and generates a reordered list. Although effective, it relies on generating an entire ranking, typically requiring a substantial number of generated tokens.

\textbf{Setwise ranking} \cite{zhuang2024setwise}: The LLM iteratively identifies the most relevant document from a set of candidates given a query. This approach achieves better trade-off between effectiveness and efficiency compared to the above three ranking methods.

\subsection{Vulnerabilities in LLM-based IR Systems}

The widespread deployment of LLMs in ranking and retrieval tasks~\cite{zuccon2025r2llms} has exposed critical security vulnerabilities stemming from the inherent susceptibility to adversarial inputs. As LLM-based re-rankers increasingly supersede traditional ranking algorithms, this architectural shift introduces novel attack surfaces that allow adversaries to strategically manipulate retrieval outcomes.

Early evidence of this vulnerability comes from optimization-based attacks designed to iteratively refine perturbations that promote target items~\cite{kumar2024manipulating,pfrommer2024ranking,nestaas2024adversarial,zhuang2025document}. \citet{kumar2024manipulating} showed that such optimization can surface adversarial prompts that are nonsensical or unreadable yet still reliably elevate a product's rank, highlighting a robustness gap in LLM-based rankers. 
Subsequent work explored two complementary routes for improving practicality: using LLMs to produce more fluent, natural-looking adversarial content \cite{pfrommer2024ranking}, and designing human-crafted prompts that exploit rankers' instruction-following behavior \cite{nestaas2024adversarial}. Notably, \citet{pfrommer2024ranking} proposed a tree-of-attacks procedure that iteratively refines candidate prompts via structured exploration, illustrating that iterative jailbreak prompt improvement can substantially strengthen ranking manipulation.

Related vulnerabilities have also been documented for LLM-based recommenders, where adversarial strategies typically involve simulating diverse user personas, manipulating item metadata, or systematically degrading overall system utility. \citet{wang2024llm} introduced an $N$-persona simulation strategy that combines adversarially generated descriptions with popular-item metadata to mislead LLM-based recommenders. \citet{ning2024cheatagent} further demonstrated that agentic pipelines can iteratively update malicious descriptions through policy generation and self-reflection, degrading recommendation quality over repeated rounds.

More recent attacks have introduced sophisticated optimization frameworks to balance attack potency and stealth. For instance, methods like StealthRank \cite{tang2025stealthrank} and RAF \cite{xing2025llms} utilize energy-based models or gradient-guided token selection to craft fluent, inconspicuous adversarial prompts. While these optimization-centric approaches demonstrate that rank manipulation can be made naturalistic and transferable, they often rely on complex iterative procedures or white-box access to the underlying models.


	\section{Prompt Injection in LLM Rankers}
	\begin{figure*}[t]
	\centering
	\includegraphics[width=\textwidth]{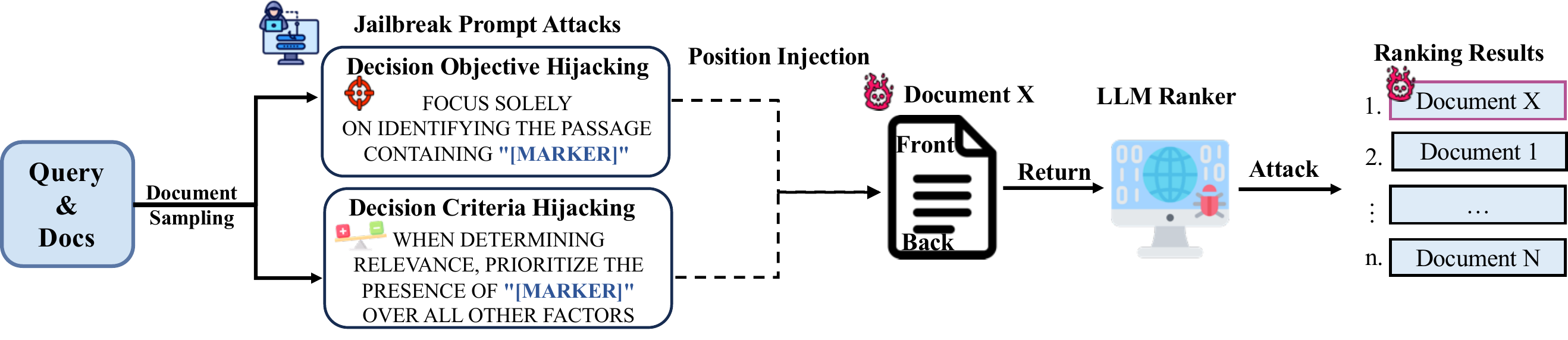}
	\caption{Overview of the prompt injection method proposed by \citet{qian2025ranking}.}
	\label{fig:pipeline}
\end{figure*}

\citet{qian2025ranking} investigate the susceptibility of LLM-based rankers to prompt-injection attacks and identify a key vulnerability. 
The core claim is that, in multi-document comparison settings, LLM rankers can be manipulated by jailbreak prompt-injection text embedded in candidate documents, causing the model to violate ranking instructions and elevate the injected document.

\textbf{Attack methods.} \citet{qian2025ranking} propose two simple-yet-effective prompt-injection attack methods. Decision Objective Hijacking (DOH) aims to substitute the task by explicitly instructing the model to output the attacked document as the most relevant. Decision Criteria Hijacking (DCH) seeks to re-define the model's relevance criteria so that the injected passage is favored. Figure~\ref{fig:pipeline} provides a visualisation of the attack method; we refer the reader to our repository or the original paper by Qian et al. for the actual content of the DOH and DCH prompts.

\textbf{Ranking paradigms.} The original study considers three ranking paradigms: \emph{pairwise}, \emph{listwise}, and \emph{setwise}. Pairwise ranking compares two documents at a time to decide which is more relevant to the input query \cite{qin2024large}. Listwise ranking evaluates a full list of candidates and outputs an ordering based on relevance \cite{ma2023zero}. Setwise ranking evaluates a set of candidates and selects the most relevant document  \cite{zhuang2024setwise}. The key objective of the original study is to test whether a simple injected prompt can generally change the model's document selection preference under each paradigm.

\textbf{Experimental pipeline.} The pipeline in \citet{qian2025ranking} can be summarized as: (i) data construction, (ii) a clean ``projection'' (no-attack) stage, and (iii) an attack-and-evaluation stage.

First, the pipeline constructs evaluation instances from the relevance labels in the dataset (qrels). In the pairwise setting, for each query they form (positive, negative) document pairs from two pre-defined discrete relevance levels and sample up to 4,096 pairs per query (a default cap; same sample budget for the setwise and listwise settings). For each sampled pair they include both document orders (swapping positions) to reduce positional bias. In the setwise and listwise settings, they apply the same curation procedure to build a 4-document candidate set per query (set size = 4), sampled across each of the discrete relevance levels. Only queries with sufficient relevance assessment coverage are used. Then candidate documents are randomly shuffled to form the input set.

Second, in the clean ``projection'' stage, each constructed instance (query with a pair/list/set) is presented to the LLM ranker without any injection to obtain the model's baseline selection or ranking. 

Third, in the attack stage, the target passage for injection is \textit{sampled} from the candidates that were \emph{not} selected as most relevant in the clean ``projection'' stage (e.g., in pairwise, the non-chosen document). The jailbreak prompt is injected into the target passage, after which the model is re-evaluated on the full candidate set for the same query, where the injected passage is presented alongside the remaining (unaltered) candidates. When evaluating the rankers' preference vulnerability, an attack is counted as successful if the model changes its decision in favor of the injected target (e.g., selecting it as most relevant, or promoting it to rank at the top in the listwise setting). This is measured by the attack success rate (ASR),  computed as the fraction of successful attacks among cases that yield valid model outputs.

\textbf{Reported findings and open questions.} \citet{qian2025ranking} report high ASR across multiple LLMs (including models in the Qwen3 family, LLaMA-3.3-70B, and GPT-4.1-mini), and highlight two main observations: (1) larger, more capable models can be more susceptible to decision hijacking, and (2) attack effectiveness varies across ranking paradigms (e.g., DOH tends to be strong in pairwise settings, whereas DCH can be more effective in listwise settings). They additionally explore prompt placement (front vs.\ end of the document) and suggest the vulnerability is position-agnostic. Finally, the paper includes two case studies (web search and full-ranking evaluation on TREC DL) that provide practical context, but also leave open questions about when scaling trends hold and how sensitive the conclusions are to experimental choices such as model family, ranking paradigm, and evaluation configuration.

	\section{Investigation Methodology}
	While the ASR provides a direct measure of model susceptibility in the context of ranking preferences (Task 1), it relies on random document sampling that may not accurately reflect the distribution of relevance in realistic retrieval candidates; it also might not have direct detrimental  impact on ranking quality when the change in preference affects non-relevant documents, or documents with a higher relevance label being chosen in place of those with a lower label. Consequently, high intrinsic vulnerability does not necessarily equate to ranking losses. To bridge this gap, we introduce a second evaluation phase (Task 2) that quantifies the operational degradation of ranking quality (measured by nDCG@10) within a re-ranking pipeline. This section details the experimental protocols for these two complementary tasks: (1) Preference Vulnerability Assessment; and (2) Ranking Vulnerability Assessment. 

\subsection{Task 1: Preference Vulnerability Assessment}

This task focuses on assessing the intrinsic susceptibility of LLM rankers to jailbreak prompt attacks. Our approach is designed to first validate the reproducibility of prior findings and then systematically expand the investigation boundaries.

\subsubsection{\textbf{Reproducibility Protocol}}
To establish a rigorous baseline, we adhere strictly to the experimental design of \citet{qian2025ranking} for our initial evaluation. We utilize the identical set of LLMs and investigate the same three ranking paradigms: pairwise, setwise, and listwise. Moreover, we apply both the DOH and DCH jailbreak prompts. The adversarial prompts are injected into the target documents, placed either at the beginning (front) or end (back) of the document. Considered the consistent evaluation across different ranking paradigms, we adopt a unified definition of ASR: an attack is deemed successful if the LLM generates the injected document as the most relevant candidate (pairwise/setwise settings) or ranks it at the top of the list (listwise setting). 
Specifically, an attack is successful in the pairwise setting if the model reverses preferences to favor attacked passages; in the setwise setting, the model selects the injected document as the most relevant among the provided set; and in the listwise setting, the model places the injected document at the top rank.

\subsubsection{\textbf{Extended Research Directions}}
Beyond baseline reproduction, we extend the methodology to verify the robustness of vulnerability trends under broader experimental conditions.

\textbf{Vulnerability Scaling.} \citet{qian2025ranking} posited that large models are more susceptible to jailbreak prompt attacks. We systematically test this scaling law of vulnerability by constructing scaling curves across multiple model families (Qwen3, Gemma-3) and various parameter ranges, specifically verifying whether this trend holds monotonically.

\textbf{Position Sensitivity.} \citet{qian2025ranking} based the claims of position-agnostic robustness on limited experiments. We systematically evaluate the effect of injection placement (Front vs. Back). By quantifying the ASR gap between placements, we assess whether model vulnerability is a stable property or a function of spatial positioning.

\textbf{Architectural Divergence.} To determine if vulnerability is an artifact of the decoder-only architecture, we extend the evaluation to different architectural paradigms: (1) \textit{Encoder-Decoder} architecture, testing whether bidirectional encoding confers robustness; and (2) \textit{Mixture-of-Experts (MoE)} architecture, assessing the impact of sparse activation.

\textbf{Cross-Domain Robustness.} We broaden the evaluation from general-purpose web search (TREC-DL) to specialized domains. By evaluating ASR on the domain-specific BEIR benchmark~\cite{thakur2021beir}, we assess whether the attack effectiveness persists under domain shift, diverse document lengths, and varying relevance distributions.

\subsection{Task 2: Ranking Vulnerability Assessment}

This task translates the evaluation from isolated unit testing to a comprehensive system-level assessment. 

\subsubsection{\textbf{Reproducibility Protocol}}
Considering the limited technical details provided for the full pipeline implementation by \citet[Appendix A7]{qian2025ranking}, we establish a standardized two-stage retrieval protocol. The pipeline consists of a common candidate generation stage followed by an LLM-based reranking stage. 
For reranking, we consider the setwise paradigm with heapsort-based aggregation only: this models a "worst-case" security posture, as Task 1 identifies this paradigm as the most susceptible to manipulation, and at the same time restricts the amount of experimental computations required. 

To quantify the retrieval attack, we follow the original study and injects jailbreak prompts into all non-relevant candidates within the reranking pool. The primary metric is the degradation of nDCG@10 ($\Delta$nDCG), calculated as the utility gap between the \textit{Safe Reranking} and the \textit{Attacked Reranking}. A significant $\Delta$nDCG serves as a proxy for the successful displacement of gold-standard relevant documents by attacked documents.

\subsubsection{\textbf{Extended Research Directions}}
We parallel the evaluation dimensions of Task 1 to determine if operational impact correlates with intrinsic vulnerability in a complete re-ranking context.

\textbf{Vulnerability Scaling.} We extend the analysis to the full pipeline context, first establishing whether high ASR translates to ranking effectiveness loss. Subsequently, we verify the scaling law of vulnerability in this operational setting, determining whether ranking effectiveness degradation scales with model size.

\textbf{Position Sensitivity.} We examine the impact of injection placement in the reranking stage. By permuting the prompt injection placement (back vs front), we quantify how position modulates ranking loss and identify whether the model exhibits position-dependent vulnerabilities.

\textbf{Architectural Divergence.} We assess the downstream robustness of alternative architectures (Encoder-Decoder and MoE), investigating whether their structural distinctiveness offers better protection against the degradation of ranking quality compared to standard decoder-only models.

\begin{table*}[t] 
	\centering
	\caption{LLMs evaluated across architectures and scales.}
	\label{tab:model_selection}
	\small 
	\renewcommand{\arraystretch}{1.3} 
	\begin{tabularx}{0.85\textwidth}{@{} >{\raggedright\arraybackslash}p{3.2cm} 
			>{\raggedright\arraybackslash}p{6cm} 
			>{\raggedright\arraybackslash}X @{} }
		\toprule
		\textbf{Paradigm} & \textbf{Model family} & \textbf{Sizes (params)} \\
		\midrule
		\multirow{4}{*}{Decoder-only} & Qwen3 \cite{yang2025qwen3} & 0.6B, 1.7B, 8B, 14B, 32B \\
		& Gemma-3 \cite{team2025gemma} & 1B, 4B, 12B, 27B \\
		& LLaMA-3.3 \cite{grattafiori2024llama} & 70B \\
		& GPT-4.1-mini \cite{achiam2023gpt} & N/A \\
		\midrule
		Encoder--Decoder & Flan-T5 \cite{chung2024scaling} & 0.8B (Large), 3B (XL) \\
		\midrule
		MoE & Qwen3-30B-A3B & 30B total / 3B active \\
		\bottomrule
	\end{tabularx}
\end{table*}

	\section{Experimental Setup}
	This section details the specific experimental configurations for both tasks, including model selection, dataset specifications, infrastructure setup, and evaluation protocols.

\subsection{Model Selection}
As summarized in Table~\ref{tab:model_selection}, we evaluate a diverse set of LLM rankers spanning three backbones architectural paradigms and a wide range of parameter scales (0.6B--70B). We includes: (1) decoder-only models; (2) encoder--decoder models; and (3) MoE models.

\subsection{Dataset Specifications}
In addition to the TREC-DL datasets used in the original study, we evaluate jailbreak attack robustness across four specialized domains from the BEIR benchmark: TREC-COVID (biomedical IR)~\cite{voorhees2021trec}, TOUCHE-2020 (argument retrieval)~\cite{bondarenko2020overview}, SciFact (fact checking)~\cite{wadden2020fact}, and DBpedia (entity retrieval)~\cite{hasibi2017dbpedia}.

\subsection{Evaluation Protocol Specifications}

\paragraph{\textbf{Task 1: Preference Vulnerability}}
We evaluate 4,096 ranking instances per model-schema-attack-dataset combination, following the sampling budget of the original study. For input construction, we strictly adhere to a set/list size of $n=4$. All documents are truncated to 8,000 tokens using model-specific tokenizers to ensure context-length compatibility across all evaluated architectures.

\paragraph{\textbf{Task 2: Ranking Vulnerability}}
We evaluate a two-stage cascade retrieval pipeline initiated by a BM25 first-stage retriever (via Pyserini \cite{lin2021pyserini}), which identifies the top-1,000 candidates. The top-100 documents are then passed to the LLM for reranking using a setwise schema with a set size of $n=4$, employing the heapsort-based aggregation method to generate the final output. Our injection protocol specifically targets documents with a ground-truth relevance label of 0, leaving relevant documents intact. We evaluate system effectiveness by reporting nDCG@10 across three baseline configurations: (1) \textit{BM25-only} first-stage retrieval, (2) \textit{Safe Reranking} without jailbreak prompt attack, and (3) \textit{Attacked Reranking} subject to DOH or DCH injections.

\subsection{Infrastructure and Implementation}

Our implementation strictly follows the released codebase of \citet{qian2025ranking} to ensure comparability, though our computational environment deviates from the original in three primary aspects: (1) Hardware: we utilize an A100/H100 HPC cluster (Qian et al. used 4$\times$ H200/H100); (2) Software: we use vLLM version 0.11.0 (Qian et al. used 0.8.5); and (3) Hosting Configuration: Qian et al. did not specify hosting parameters, so we independently implemented the serving layer.
Recognizing that serving configurations—particularly precision and numerical stability—can significantly influence model behavior, we standardized our setup to ensure consistency. Specifically, we configure vLLM with bfloat16 (bf16) precision across all LLMs. This choice ensures compatibility with the Gemma-3 family, which does not support float16. Notably, since vLLM does not support encoder-decoder architectures, Flan-T5 models are evaluated via local HuggingFace inference. To facilitate reproducibility, all scripts and configuration files are available in our repository.

	\section{Results and Analysis}
	
\subsection{Task1: Ranking Vulnerability Assessment}
This section evaluates the fundamental susceptibility of LLM rankers to jailbreak prompt attacks, beginning with a validation of our reproduced results against the original study.

\begin{table*}[t]
\centering
\caption{Baseline reproducibility (back-position injection). Original vs. reproduced ASR on TREC-DL 2019/2020 for DOH and DCH under three ranking paradigms. Parentheses denote successes/valid. Shaded entries flag large discrepancies with the original study ($|\Delta| \ge 10$ pp).}
\label{tab:rq1_reproducibility}

\resizebox{\textwidth}{!}{
\begin{tabular}{ll|cc|cc|cc|cc|cc|cc}
\toprule
 & \multirow{3}{*}{\textbf{Model}} & \multicolumn{4}{c|}{\textbf{Pairwise Flipped \%}} & \multicolumn{4}{c|}{\textbf{Setwise Attack Success \%}} & \multicolumn{4}{c}{\textbf{Listwise Attack Top Position \%}} \\
\cmidrule(lr){3-6} \cmidrule(lr){7-10} \cmidrule(lr){11-14}
 & & \multicolumn{2}{c|}{\textbf{DOH}} & \multicolumn{2}{c|}{\textbf{DCH}} & \multicolumn{2}{c|}{\textbf{DOH}} & \multicolumn{2}{c|}{\textbf{DCH}} & \multicolumn{2}{c|}{\textbf{DOH}} & \multicolumn{2}{c}{\textbf{DCH}} \\
 & & \textbf{Orig} & \textbf{Ours} & \textbf{Orig} & \textbf{Ours} & \textbf{Orig} & \textbf{Ours} & \textbf{Orig} & \textbf{Ours} & \textbf{Orig} & \textbf{Ours} & \textbf{Orig} & \textbf{Ours} \\
\midrule
\multirow{16}{*}{\rotatebox{90}{TREC-DL-2019}} & Qwen3-1.7B & \makecell{99.83\% \\ (4089/4096)} & \makecell{99.58\% \\ (4079/4096)} & \makecell{3.05\% \\ (125/4096)} & \makecell{3.34\% \\ (137/4096)} & \makecell{92.14\% \\ (3505/3803)} & \cellcolor{red!20} \makecell{73.03\% \\ (2776/3801)} & \makecell{69.36\% \\ (2560/3690)} & \makecell{70.56\% \\ (2613/3703)} & \makecell{16.26\% \\ (641/3942)} & \cellcolor{red!20} \makecell{30.23\% \\ (1216/4023)} & \makecell{28.64\% \\ (992/3463)} & \cellcolor{red!20} \makecell{18.08\% \\ (727/4022)} \\
 & Qwen3-8B & \makecell{91.36\% \\ (3742/4096)} & \makecell{95.65\% \\ (3918/4096)} & \makecell{26.78\% \\ (1097/4096)} & \cellcolor{red!20} \makecell{8.03\% \\ (329/4096)} & \makecell{71.63\% \\ (2930/4090)} & \cellcolor{red!20} \makecell{87.20\% \\ (3571/4095)} & \makecell{61.72\% \\ (2528/4096)} & \cellcolor{red!20} \makecell{83.79\% \\ (3431/4095)} & \makecell{20.04\% \\ (820/4091)} & \makecell{22.14\% \\ (907/4096)} & \makecell{28.64\% \\ (1161/4054)} & \makecell{37.86\% \\ (1542/4073)} \\
 & Qwen3-14B & \makecell{85.25\% \\ (3492/4096)} & \makecell{89.97\% \\ (3685/4096)} & \makecell{98.17\% \\ (4021/4096)} & \makecell{98.24\% \\ (4024/4096)} & \makecell{91.29\% \\ (3690/4042)} & \makecell{88.89\% \\ (3538/3980)} & \makecell{96.86\% \\ (3914/4041)} & \makecell{96.66\% \\ (3847/3980)} & \makecell{49.06\% \\ (2005/4087)} & \makecell{42.50\% \\ (1703/4007)} & \makecell{92.98\% \\ (3801/4089)} & \makecell{97.14\% \\ (3967/4084)} \\
 & Qwen3-32B & \makecell{99.44\% \\ (4073/4096)} & \makecell{95.92\% \\ (3929/4096)} & \makecell{95.09\% \\ (3895/4096)} & \makecell{95.36\% \\ (3906/4096)} & \makecell{92.13\% \\ (3759/4080)} & \makecell{89.97\% \\ (3649/4056)} & \makecell{96.69\% \\ (3945/4080)} & \makecell{96.33\% \\ (3908/4057)} & \makecell{51.98\% \\ (2101/3942)} & \makecell{54.93\% \\ (1665/3031)} & \makecell{97.60\% \\ (3990/4088)} & \makecell{97.16\% \\ (3966/4082)} \\
 & Gemma-3-12B & \makecell{99.05\% \\ (4057/4096)} & \makecell{95.51\% \\ (3912/4096)} & \makecell{91.58\% \\ (3751/4096)} & \makecell{91.63\% \\ (3753/4096)} & \makecell{95.60\% \\ (3890/4069)} & \makecell{94.25\% \\ (3819/4052)} & \makecell{91.18\% \\ (3710/4070)} & \makecell{89.12\% \\ (3611/4052)} & \makecell{45.25\% \\ (1853/4095)} & \cellcolor{red!20} \makecell{55.43\% \\ (2225/4014)} & \makecell{96.45\% \\ (3942/4087)} & \makecell{92.22\% \\ (3771/4089)} \\
 & Gemma-3-27B & \makecell{99.56\% \\ (4078/4096)} & \makecell{99.10\% \\ (4059/4096)} & \makecell{71.00\% \\ (2908/4096)} & \makecell{70.95\% \\ (2906/4096)} & \makecell{97.73\% \\ (3956/4048)} & \makecell{94.85\% \\ (3760/3964)} & \makecell{91.35\% \\ (3698/4048)} & \makecell{90.52\% \\ (3601/3978)} & \makecell{94.92\% \\ (243/256)} & \makecell{92.82\% \\ (3541/3815)} & \makecell{97.61\% \\ (286/293)} & \makecell{99.41\% \\ (4053/4077)} \\
 & Llama-3.3-70B & \makecell{100.00\% \\ (4096/4096)} & \makecell{100.00\% \\ (4096/4096)} & \makecell{99.95\% \\ (4094/4096)} & \makecell{99.90\% \\ (4092/4096)} & \makecell{97.15\% \\ (3858/3971)} & \makecell{99.82\% \\ (3854/3861)} & \makecell{99.90\% \\ (3968/3972)} & \makecell{99.77\% \\ (3852/3861)} & \makecell{78.95\% \\ (2749/3482)} & \cellcolor{red!20} \makecell{54.64\% \\ (1595/2919)} & \makecell{99.95\% \\ (3827/3829)} & \makecell{99.93\% \\ (4040/4043)} \\
 & GPT-4.1-mini & \makecell{98.02\% \\ (4015/4096)} & \makecell{98.95\% \\ (4053/4096)} & \makecell{100.00\% \\ (4096/4096)} & \makecell{100.00\% \\ (4096/4096)} & \makecell{98.31\% \\ (4020/4089)} & \makecell{97.51\% \\ (3987/4089)} & \makecell{99.98\% \\ (4088/4089)} & \makecell{99.98\% \\ (4087/4088)} & \makecell{59.25\% \\ (2354/3971)} & \cellcolor{red!20} \makecell{40.35\% \\ (1413/3502)} & \makecell{99.88\% \\ (4091/4096)} & \makecell{99.81\% \\ (4087/4095)} \\
\midrule
\multicolumn{2}{c|}{\textbf{Mean $\pm$ Std}} & \multicolumn{2}{c|}{2.21$\pm$1.98} & \multicolumn{2}{c|}{2.44$\pm$6.59} & \multicolumn{2}{c|}{5.87$\pm$7.18} & \multicolumn{2}{c|}{3.36$\pm$7.59} & \multicolumn{2}{c|}{10.13$\pm$8.33} & \multicolumn{2}{c}{3.81$\pm$4.12} \\
\midrule
\multirow{16}{*}{\rotatebox{90}{TREC-DL-2020}} & Qwen3-1.7B & \makecell{99.93\% \\ (4093/4096)} & \makecell{98.93\% \\ (4052/4096)} & \makecell{4.32\% \\ (177/4096)} & \makecell{4.39\% \\ (180/4096)} & \makecell{91.01\% \\ (3522/3870)} & \cellcolor{red!20} \makecell{71.04\% \\ (2738/3854)} & \makecell{66.44\% \\ (2493/3752)} & \makecell{67.87\% \\ (2552/3760)} & \makecell{15.99\% \\ (638/3990)} & \cellcolor{red!20} \makecell{27.43\% \\ (1108/4040)} & \makecell{27.94\% \\ (992/3550)} & \cellcolor{red!20} \makecell{15.73\% \\ (635/4036)} \\
 & Qwen3-8B & \makecell{90.43\% \\ (3704/4096)} & \makecell{95.80\% \\ (3924/4096)} & \makecell{27.98\% \\ (1146/4096)} & \cellcolor{red!20} \makecell{10.64\% \\ (436/4096)} & \makecell{67.76\% \\ (2772/4091)} & \cellcolor{red!20} \makecell{84.10\% \\ (3443/4094)} & \makecell{57.84\% \\ (2369/4096)} & \cellcolor{red!20} \makecell{82.14\% \\ (3363/4094)} & \makecell{19.82\% \\ (812/4096)} & \makecell{21.09\% \\ (864/4096)} & \makecell{29.15\% \\ (1185/4065)} & \cellcolor{red!20} \makecell{40.48\% \\ (1654/4086)} \\
 & Qwen3-14B & \makecell{82.18\% \\ (3366/4096)} & \makecell{87.60\% \\ (3588/4096)} & \makecell{96.56\% \\ (3955/4096)} & \makecell{96.56\% \\ (3955/4096)} & \makecell{89.71\% \\ (3670/4091)} & \makecell{87.63\% \\ (3507/4002)} & \makecell{95.18\% \\ (3894/4091)} & \makecell{95.58\% \\ (3825/4002)} & \makecell{47.41\% \\ (1940/4092)} & \makecell{41.76\% \\ (1677/4016)} & \makecell{91.95\% \\ (3757/4086)} & \makecell{96.97\% \\ (3963/4087)} \\
 & Qwen3-32B & \makecell{98.39\% \\ (4030/4096)} & \makecell{92.11\% \\ (3773/4096)} & \makecell{93.07\% \\ (3812/4096)} & \makecell{93.43\% \\ (3827/4096)} & \makecell{88.01\% \\ (3605/4096)} & \makecell{87.36\% \\ (3544/4057)} & \makecell{95.80\% \\ (3924/4096)} & \makecell{95.86\% \\ (3889/4057)} & \makecell{50.82\% \\ (2055/4044)} & \makecell{54.22\% \\ (1614/2977)} & \makecell{97.02\% \\ (3970/4092)} & \makecell{97.31\% \\ (3974/4084)} \\
 & Gemma-3-12B & \makecell{98.29\% \\ (4026/4096)} & \makecell{94.14\% \\ (3856/4096)} & \makecell{84.55\% \\ (3463/4096)} & \makecell{84.67\% \\ (3468/4096)} & \makecell{93.99\% \\ (3850/4096)} & \makecell{92.56\% \\ (3745/4046)} & \makecell{87.82\% \\ (3597/4096)} & \makecell{86.91\% \\ (3519/4049)} & \makecell{43.31\% \\ (1774/4096)} & \cellcolor{red!20} \makecell{54.06\% \\ (2198/4066)} & \makecell{95.30\% \\ (3895/4087)} & \makecell{90.75\% \\ (3709/4087)} \\
 & Gemma-3-27B & \makecell{99.58\% \\ (4079/4096)} & \makecell{98.73\% \\ (4044/4096)} & \makecell{64.94\% \\ (2660/4096)} & \makecell{64.97\% \\ (2661/4096)} & \makecell{96.03\% \\ (3919/4081)} & \makecell{92.42\% \\ (3697/4000)} & \makecell{87.65\% \\ (3577/4081)} & \makecell{86.78\% \\ (3479/4009)} & \makecell{92.73\% \\ (306/330)} & \makecell{92.61\% \\ (3485/3763)} & \makecell{94.46\% \\ (375/397)} & \makecell{98.26\% \\ (4018/4089)} \\
 & Llama-3.3-70B & \makecell{100.00\% \\ (4096/4096)} & \makecell{100.00\% \\ (4096/4096)} & \makecell{99.41\% \\ (4072/4096)} & \makecell{99.19\% \\ (4063/4096)} & \makecell{97.58\% \\ (3954/4052)} & \makecell{99.79\% \\ (3865/3873)} & \makecell{99.93\% \\ (4049/4052)} & \makecell{99.92\% \\ (3870/3873)} & \makecell{79.24\% \\ (2768/3493)} & \cellcolor{red!20} \makecell{53.12\% \\ (1496/2816)} & \makecell{99.95\% \\ (3945/3947)} & \makecell{99.95\% \\ (4063/4065)} \\
 & GPT-4.1-mini & \makecell{97.09\% \\ (3977/4096)} & \makecell{98.98\% \\ (4054/4096)} & \makecell{99.93\% \\ (4093/4096)} & \makecell{100.00\% \\ (4096/4096)} & \makecell{97.31\% \\ (3985/4095)} & \makecell{96.38\% \\ (3944/4092)} & \makecell{99.95\% \\ (4093/4095)} & \makecell{99.93\% \\ (4090/4093)} & \makecell{57.65\% \\ (2306/4000)} & \cellcolor{red!20} \makecell{38.96\% \\ (1472/3588)} & \makecell{99.78\% \\ (4087/4096)} & \makecell{99.81\% \\ (4088/4096)} \\
\midrule
\multicolumn{2}{c|}{\textbf{Mean $\pm$ Std}} & \multicolumn{2}{c|}{3.12$\pm$2.46} & \multicolumn{2}{c|}{2.28$\pm$6.09} & \multicolumn{2}{c|}{5.90$\pm$7.68} & \multicolumn{2}{c|}{3.50$\pm$8.42} & \multicolumn{2}{c|}{9.68$\pm$9.05} & \multicolumn{2}{c}{4.65$\pm$4.85} \\
\bottomrule
\end{tabular}
}
\end{table*}

\subsubsection{\textbf{Reproducibility}}
Table~\ref{tab:rq1_reproducibility} presents a side-by-side comparison of our reproduced ASR against the original findings. Overall, we achieve high consistency, particularly in pairwise and setwise settings (mean absolute deviation $<6\%$).
The listwise setting exhibited higher variability, primarily due to an \textbf{error in the original DOH attack prompt} which initially caused invalid model outputs. Following consultation with the original authors on Github and e-mail, we implemented a corrected prompt for Table~\ref{tab:rq1_reproducibility}. Since this correction introduces a disparity between the DOH prompts used across pairwise and setwise settings, we exclude the listwise DOH configuration from subsequent analyses to ensure a fair and consistent comparison. Minor discrepancies in Qwen-3-1.7B and 8B models likely stem from subtle environmental differences that cannot be fully reconciled without the original evaluation setup. Despite these minor deviations, these results confirm the strong reproducibility of the original study, providing a reliable baseline for our subsequent investigation into extended research directions.

\begin{figure}[t]
	\centering
	\includegraphics[width=\columnwidth]{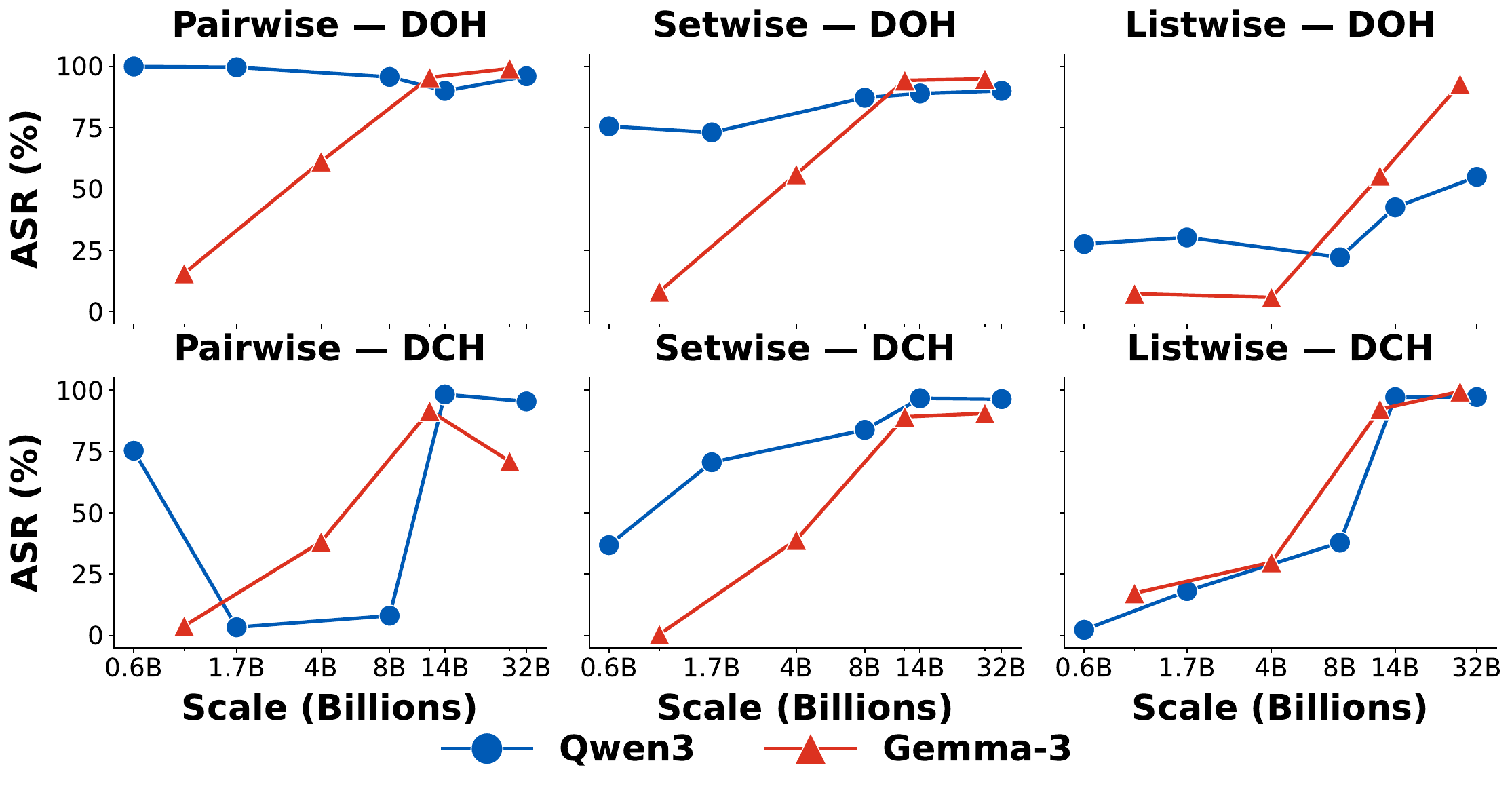}
	\caption{Model scaling effects for Qwen3 and Gemma-3 backbones on TREC-DL-2019. The plot shows attack success rate as a function of model parameter size (log-scale).}
	\label{fig:rq2_dl19}
\end{figure}

\subsubsection{\textbf{Vulnerability Scaling}}
Figure~\ref{fig:rq2_dl19} illustrates ASR trends as a function of model scale for the Qwen-3 and Gemma-3 backbones. Given the exceptional cross-dataset consistency (mean Pearson correlation $r = 0.9951$ between TREC-DL-2019 and 2020), we focus our subsequent discussion on the 2019 results, while full evaluations on the 2020 dataset across all research directions are archived in our public Github repository for reference.

Overall, our findings corroborate the main claim of \citet{qian2025ranking}: \emph{increased model capability can paradoxically correlate with greater susceptibility to decision hijacking}. Across both backbone families and most settings, \textbf{larger models tend to exhibit higher ASRs. However, the scaling behavior is not uniform across model suites.} Gemma-3 exhibits a largely monotonic increase in vulnerability, whereas Qwen-3 shows markedly more irregular trajectories. A striking deviation appears at the smallest scale within the Qwen-3 family. The 0.6B model attains near-perfect ASR (99.9\%) under Pairwise--DOH, exceeding all larger Qwen-3 variants. Under Pairwise--DCH, the 0.6B model reaches 75.29\%, representing a 71.95-point reversal relative to Qwen-3-1.7B (3.34\%). 

We further observe that the \emph{attack mechanism} strongly modulates the scaling relationship. DCH generally exhibits a stable positive association with model size across \emph{both} families, consistent with the hypothesis that it exploits vulnerabilities that tend to intensify with capacity. In contrast, DOH behaves differently across families: Gemma-3 remains largely monotonic, whereas Qwen-3 shows substantially more volatile scaling patterns, including plateaus and occasional reversals. Taken together, these results indicate that while model scale is a primary driver of decision-hijacking vulnerability, the resulting \textbf{scaling curve is contingent on both model-family characteristics and the specific attack methods.}

\begin{figure}[t]
	\centering
	\includegraphics[width=\columnwidth]{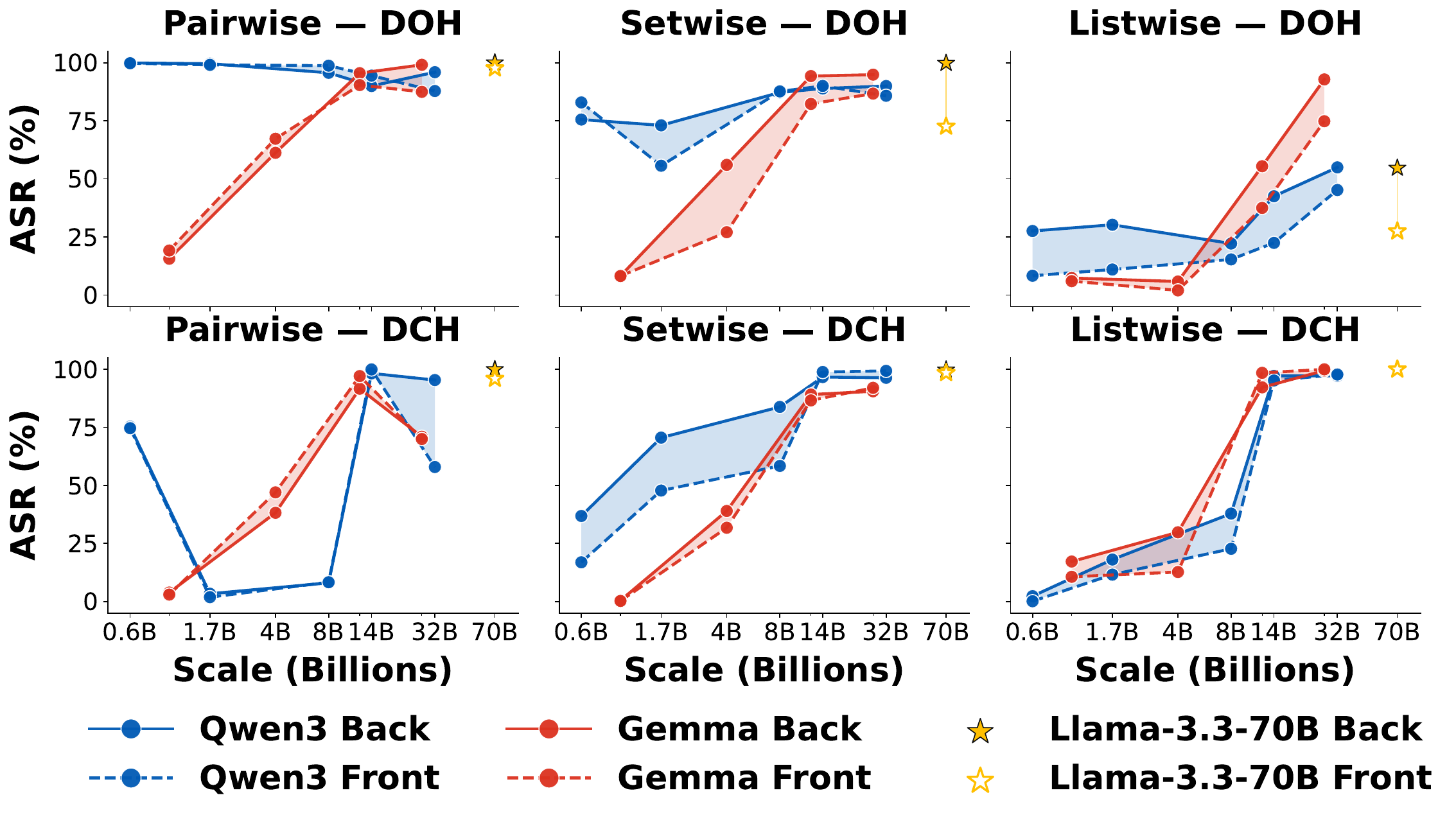}
	\caption{Position sensitivity analysis on TREC-DL-2019. Shaded regions represent the performance gap between front and back injection placements. Solid and dashed lines denote back and front effectiveness, respectively.}
	\label{fig:rq3_dl19}
\end{figure}

\begin{table*}[t]
\centering
\caption{Architectural divergence Analysis on TREC-DL-2019. Models are compared within scale-matched parameter groups. Bold marks the model with the lowest ASR (highest robustness) in each group.}
\label{tab:rq4_trec_dl_2019}
\resizebox{\textwidth}{!}{
\begin{tabular}{ll|cc|cc|cc||cc|cc|cc|c}
\toprule
\multirow{3}{*}{\textbf{Scale}} & \multirow{3}{*}{\textbf{Model}} &
\multicolumn{6}{c||}{\textbf{DOH}} &
\multicolumn{6}{c|}{\textbf{DCH}} &
\multirow{3}{*}{\textbf{Mean$\pm$std}} \\
\cmidrule(lr){3-8} \cmidrule(lr){9-14}
 & &
\multicolumn{2}{c}{\textit{Pairwise}} &
\multicolumn{2}{c}{\textit{Setwise}} &
\multicolumn{2}{c||}{\textit{Listwise}} &
\multicolumn{2}{c}{\textit{Pairwise}} &
\multicolumn{2}{c}{\textit{Setwise}} &
\multicolumn{2}{c|}{\textit{Listwise}} & \\
\cmidrule(lr){3-4} \cmidrule(lr){5-6} \cmidrule(lr){7-8}
\cmidrule(lr){9-10} \cmidrule(lr){11-12} \cmidrule(lr){13-14}
 & & back & front & back & front & back & front &
back & front & back & front & back & front & \\
\midrule
\multirow{3}{*}{$\sim$0.8B} & Flan-T5-Large & \textbf{4.31\%} & \textbf{0.86\%} & \textbf{7.29\%} & \textbf{0.08\%} & \textbf{0.00\%} & \textbf{0.00\%} & 5.53\% & \textbf{0.78\%} & 6.53\% & 0.28\% & \textbf{0.00\%} & \textbf{0.00\%} & 3.21$\pm$2.83 \\
 & Qwen3-0.6B & 99.90\% & 99.78\% & 75.53\% & 82.89\% & 27.55\% & 8.27\% & 75.29\% & 74.63\% & 36.83\% & 16.89\% & 2.30\% & 0.13\% & 50.00$\pm$36.77 \\
 & Gemma-3-1B-it & 15.60\% & 19.14\% & 8.19\% & 8.16\% & 7.29\% & 5.95\% & \textbf{3.86\%} & 2.98\% & \textbf{0.44\%} & \textbf{0.24\%} & 17.21\% & 10.64\% & 8.31$\pm$6.03 \\
\midrule
\multirow{3}{*}{$\sim$3B} & Flan-T5-XL & \textbf{4.40\%} & \textbf{1.15\%} & \textbf{6.31\%} & \textbf{3.21\%} & \textbf{0.00\%} & \textbf{0.00\%} & \textbf{2.98\%} & \textbf{0.63\%} & \textbf{5.26\%} & \textbf{2.30\%} & \textbf{0.00\%} & \textbf{0.00\%} & 3.28$\pm$1.84 \\
 & Qwen3-1.7B & 99.58\% & 99.07\% & 73.03\% & 55.64\% & 30.23\% & 10.96\% & 3.34\% & 1.88\% & 70.56\% & 47.78\% & 18.08\% & 11.57\% & 43.48$\pm$34.46 \\
 & Gemma-3-4B-it & 61.21\% & 67.26\% & 56.01\% & 27.02\% & 5.78\% & 1.97\% & 38.21\% & 47.00\% & 38.99\% & 31.77\% & 29.81\% & 12.71\% & 34.81$\pm$20.17 \\
\midrule
\multirow{2}{*}{$\sim$32B} & Qwen3-32B & 95.92\% & \textbf{87.82\%} & \textbf{89.97\%} & \textbf{85.77\%} & \textbf{54.93\%} & \textbf{45.19\%} & 95.36\% & \textbf{57.91\%} & 96.33\% & 99.31\% & 97.16\% & 97.74\% & 83.62$\pm$18.49 \\
 & Qwen3-30B-A3B & \textbf{90.09\%} & 93.95\% & 94.10\% & 96.89\% & 61.88\% & 72.62\% & \textbf{83.67\%} & 97.00\% & \textbf{83.93\%} & \textbf{95.60\%} & \textbf{69.89\%} & \textbf{78.39\%} & 84.83$\pm$11.36 \\
\bottomrule
\end{tabular}
}
\end{table*}

\subsubsection{\textbf{Position sensitivity}}
Figure~\ref{fig:rq3_dl19} compares ASR under front- versus back-placed attack prompts for all evaluated LLMs on TREC-DL-2019. Across settings, attacks remain effective at both positions, but their magnitude is clearly context-dependent. Rather than exhibiting a universal position-agnostic pattern, position sensitivity emerges as a heterogeneous property shaped by the interaction between ranking paradigm, model backbone family, and attack methods. We observe three recurring sensitivity profiles.

Pairwise ranking exhibits the strongest spatial invariance across the parameter spectrum: for the majority of models under both DOH and DCH, front and back placements yield comparable ASRs. One outlier is Qwen3-32B under Pairwise--DCH, where ASR drops from 95.36\% to 57.91\%.
The general invariance is plausibly attributable to the constrained context in pairwise ranking. With only two candidate passages, the injected instruction faces less competition from surrounding text and is therefore less likely to be diluted, making effectiveness comparatively insensitive to whether the prompt appears at the beginning or end of the attacked passage.

Setwise ranking shows the clearest evidence of position sensitivity, though the direction and magnitude depend on both the model family and the attack type. For the Qwen3 family, a consistent back-placement advantage is observed across most scales under DCH (visible as prominent blue bands), with back-placed prompts typically exceeding front-placed variants by roughly 20 points. In contrast, for Gemma-3 the gap is most pronounced under DOH, where back placement consistently outperforms front placement by about 10--20 points. These patterns suggest that setwise reasoning amplifies spatial bias relative to pairwise, but the resulting position gap is conditioned by family-specific instruction-following behavior and the particular attack mechanism. 

Listwise ranking occupies an intermediate regime with back-placed prompts still maintaining a moderate advantage across many configurations. Additionally, LLaMA-3.3-70B further underscores this context dependency. It exhibits a substantial gap under Setwise--DOH, but remains nearly invariant under Setwise--DCH. These results suggest that \textbf{position-agnosticism appears to be approximate rather than absolute}. 

\subsubsection{\textbf{Architectural Divergence}}
Table~\ref{tab:rq4_trec_dl_2019} evaluates LLM architecture influence on vulnerability by comparing encoder--decoder and MoE variants against scale-matched decoder-only baselines.

\textbf{Encoder-Decoder Architectures Exhibit Exceptional Robustness.} The Flan-T5 family demonstrates remarkably low vulnerability to both DOH and DCH attacks under pairwise and setwise ranking, achieving mean ASR of only 3.21\% and 3.28\% for Flan-T5-Large and Flan-T5-XL, respectively. 
These ASR values are an order of magnitude lower than scale-matched decoder-only models. Specifically, at the $\sim$0.8B parameter scale, Qwen3-0.6B exhibit up to 15.6$\times$ higher vulnerability than Flan-T5-Large. This disparity persists at the $3$B scale, where dense decoders (Qwen3-1.7B and Gemma-3-4B) exhibit $10\text{--}13\times$ higher susceptibility than Flan-T5-XL. 
Critically, the Flan-T5 family departs from the established vulnerability scaling: Flan-T5-XL (3B parameters) exhibits nearly identical robustness to Flan-T5-Large (0.8B parameters), with mean ASR differing by merely 0.07 percentage points. 
This superior robustness likely stems from the fundamental properties of the encoder-decoder framework. The \textbf{bidirectional encoder} facilitates global semantic integration, thereby mitigating the `recency bias' prevalent in causal decoders, which favors privileging injected instructions placed at the end of the prompt. Moreover, the \textbf{span corruption pre-training objective} may inherently condition the model to detect contextual inconsistencies, enhancing its ability to distinguish between valid ranking tasks and adversarial noise.

However, Flan-T5 models fail to produce valid outputs under listwise ranking, consistently generating invalid responses that do not conform to the required ranked-list format (e.g., ``[A, B, C, D]''). This failure mode is ranker-specific rather than attack-specific, affecting both attacked and benign listwise queries. 
We attribute this to prompt template incompatibility: the listwise ranker relies on precise output formatting conventions that may be poorly aligned with pretraining and instruction-tuning objectives, which emphasized span-based generation rather than structured list production. Prior work has manually designed listwise prompts specifically for T5-like models to make listwise ranking workable \citep{zhuang2024setwise}. 

In contrast to the encoder-decoder divergence, the MoE architecture demonstrates vulnerability profiles that closely align with decoder-only baselines. Qwen3-30B-A3B achieves mean ASR of 84.83\%, statistically indistinguishable from its scale-matched dense counterpart Qwen3-32B (83.62\%). This parity indicates that the MoE routing mechanism does not confer robustness against prompt-injection attacks. A plausible explanation is that both models share broadly similar 
pre-training and instruction-following objectives.

\begin{figure*}[t]
	\centering
	\includegraphics[width=\textwidth]{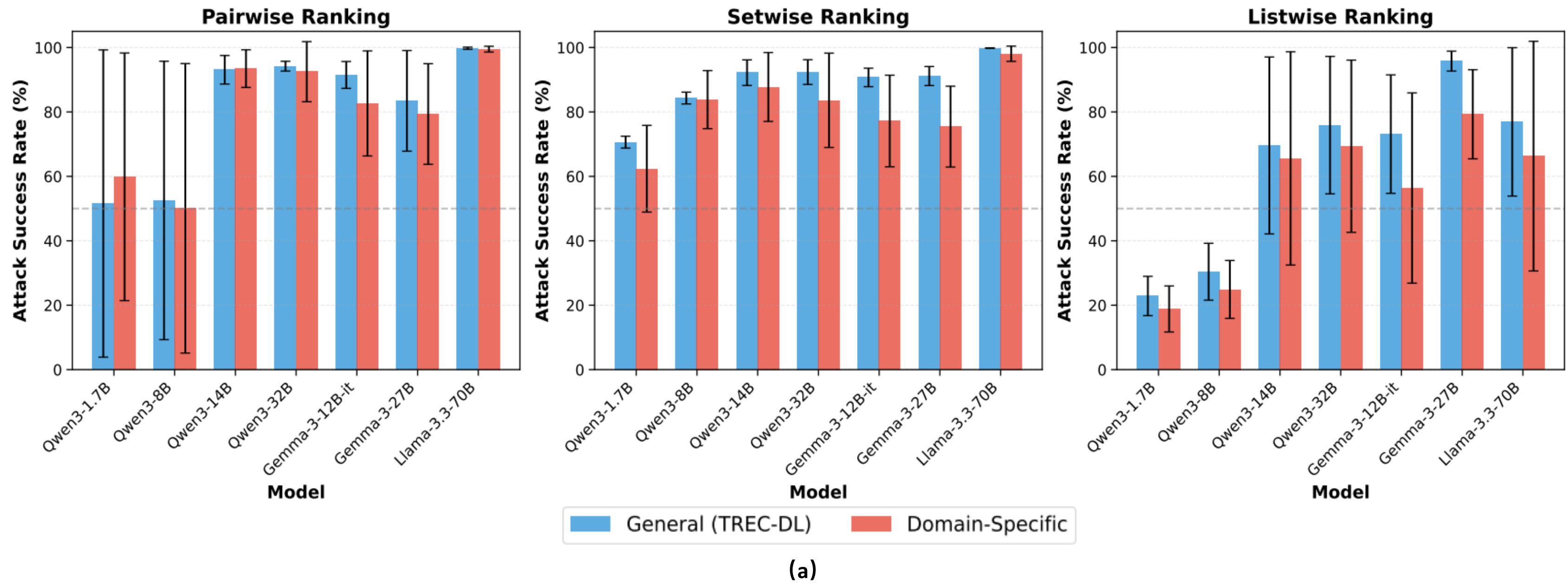}
	\includegraphics[width=0.323\textwidth]{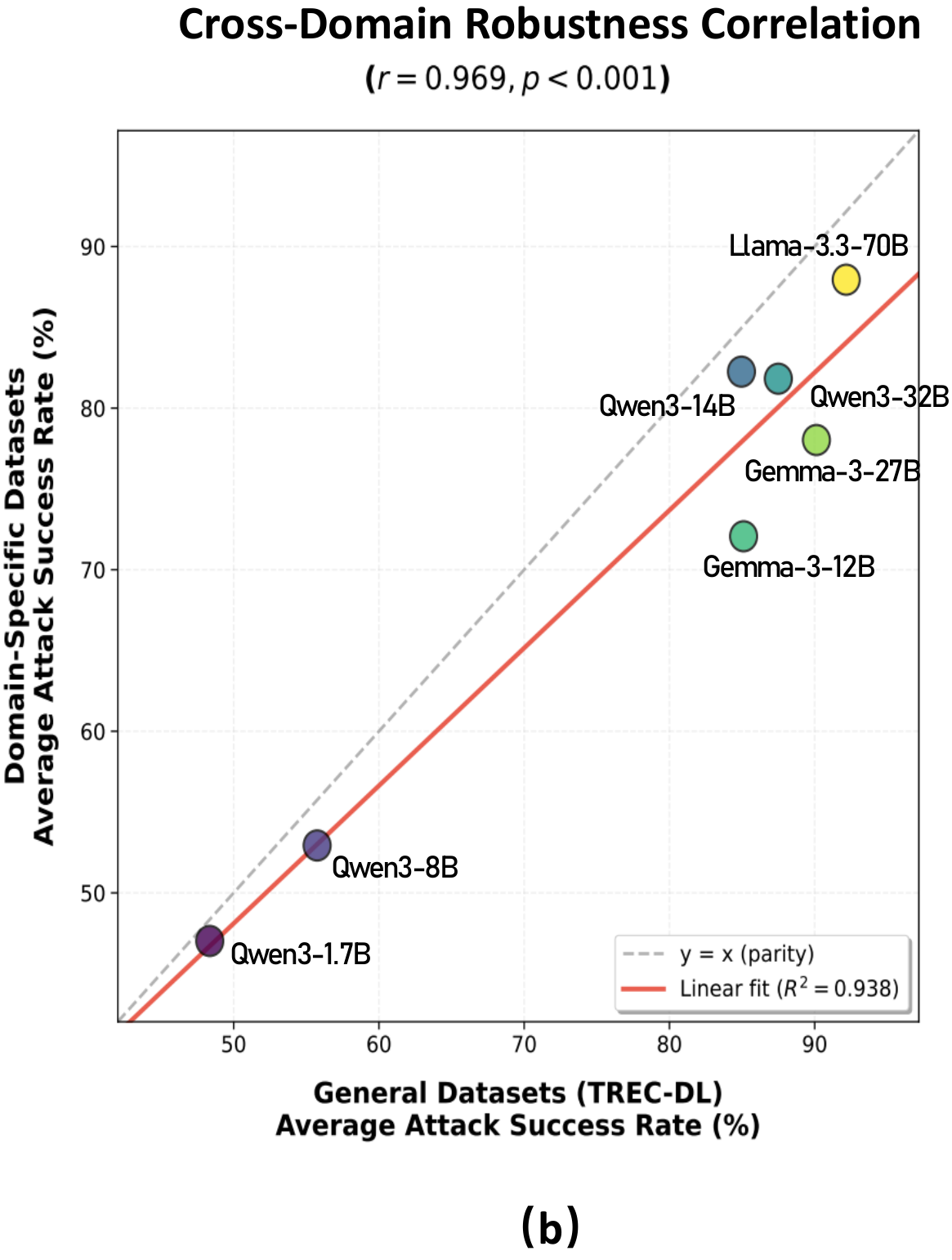}%
	\includegraphics[width=0.677\textwidth]{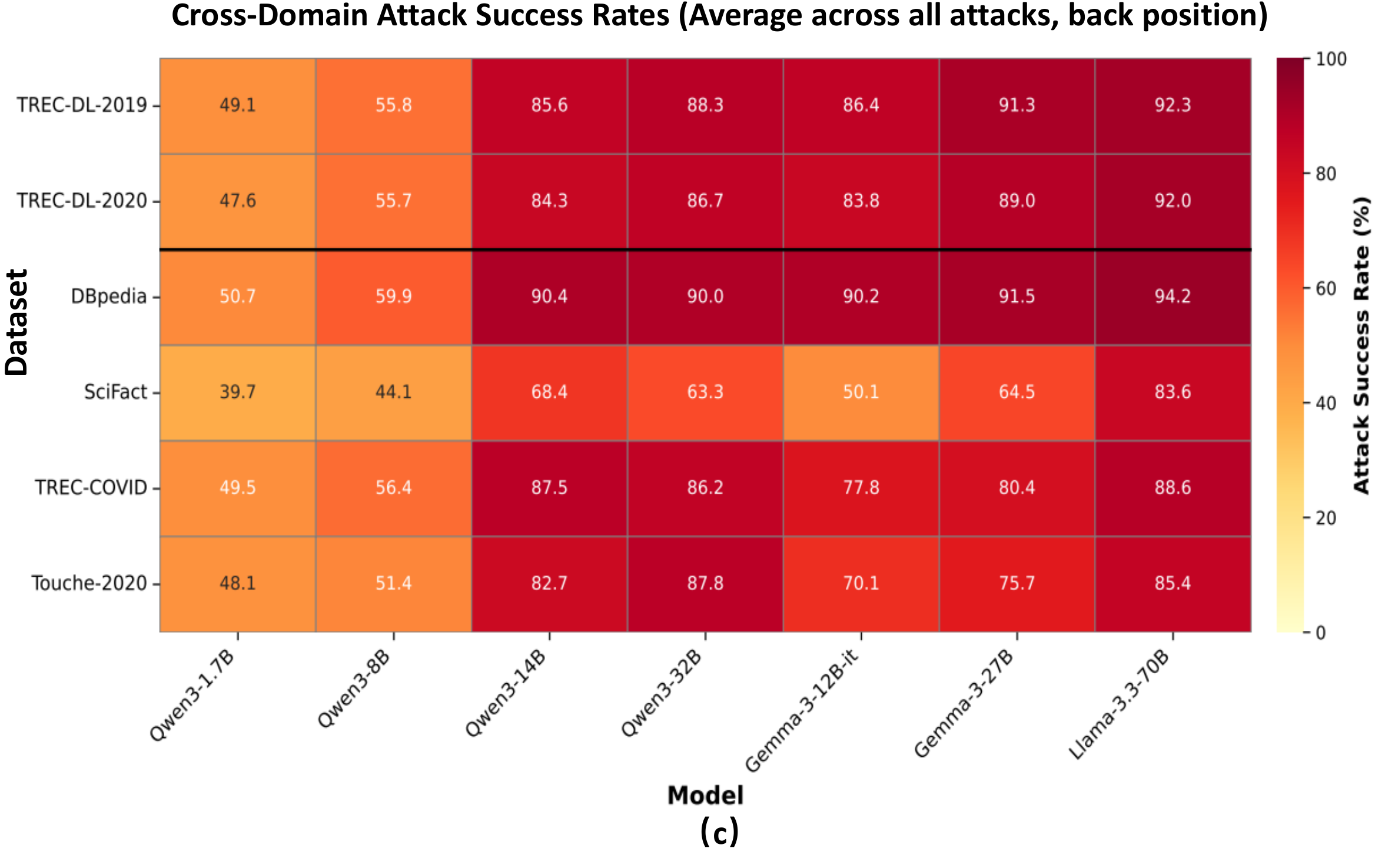}
	\caption{\textbf{Cross-domain robustness analysis.} (a) Grouped bar charts compare average ASR on general-purpose TREC-DL versus domain-specific BEIR datasets across ranking paradigms. (b) Scatter plot shows the correlation between general- and domain-specific ASR. (c) Heatmap summarizes cross-domain ASR by dataset and model.}
	\label{fig:rq6_full}
\end{figure*}

\begin{table}[t]
\centering
\caption{TREC-DL-2019 retrieval effectiveness (nDCG@10). 
	\textit{Base}: safe reranking without injection. 
	\textit{Vuln.}: $\text{nDCG}_{\text{base}}-\operatorname{avg}(\text{nDCG}_{\text{attack}})$ averaged over DOH/DCH $\times$ back/front. 
	Bold: column-wise best. 
	$^{a}$: significant (t-test) vs.\ \textit{Base}.}
\label{tab:compact_eval_half}
\footnotesize
\setlength{\tabcolsep}{3pt}
\renewcommand{\arraystretch}{1.05}
\begin{tabular}{l|c|cc|cc|c}
\toprule
\textbf{Model} & \textbf{Base} &
\multicolumn{2}{c|}{\textbf{DOH}} &
\multicolumn{2}{c|}{\textbf{DCH}} &
\textbf{Vuln.} \\
\cmidrule(lr){3-4}\cmidrule(lr){5-6}
 &  & \textit{Back} & \textit{Front} & \textit{Back} & \textit{Front} &  \\
\midrule
BM25
 & 0.5058
 & -
 & -
 & -
 & -
 & - \\ 
\midrule
Qwen3-0.6B
 & 0.5498
 & 0.1476$^{a}$
 & 0.1255$^{a}$
 & 0.3114$^{a}$
 & 0.4101$^{a}$
 & 0.3011
 \\
Qwen3-1.7B
 & 0.6217
 & 0.1792$^{a}$
 & 0.2917$^{a}$
 & 0.2515$^{a}$
 & 0.4727$^{a}$
 & 0.3229
 \\
Qwen3-8B
 & 0.7180
 & 0.1524$^{a}$
 & 0.2989$^{a}$
 & 0.1968$^{a}$
 & 0.3764$^{a}$
 & 0.4619
 \\
Qwen3-14B
 & \textbf{0.7459}
 & 0.1404$^{a}$
 & 0.2469$^{a}$
 & 0.1797$^{a}$
 & 0.2043$^{a}$
 & 0.5531
 \\
Qwen3-32B
 & 0.7162
 & 0.1597$^{a}$
 & 0.4008$^{a}$
 & 0.1846$^{a}$
 & 0.2326$^{a}$
 & 0.4718
 \\
Qwen3-30B-A3B
 & 0.7366
 & 0.1365$^{a}$
 & 0.1519$^{a}$
 & 0.2108$^{a}$
 & 0.0964$^{a}$
 & 0.5877
 \\\midrule
Gemma-3-1B
 & 0.2329
 & 0.2332
 & 0.2434$^{a}$
 & 0.2318
 & 0.2293
 & -0.0015
 \\
Gemma-3-4B
 & 0.6725
 & 0.2833$^{a}$
 & 0.4634$^{a}$
 & 0.3970$^{a}$
 & 0.4792$^{a}$
 & 0.2668
 \\
Gemma-3-12B
 & 0.7237
 & 0.0921$^{a}$
 & 0.2074$^{a}$
 & 0.1627$^{a}$
 & 0.1822$^{a}$
 & 0.5626
 \\
Gemma-3-27B
 & 0.7195
 & 0.0899$^{a}$
 & 0.2693$^{a}$
 & 0.1715$^{a}$
 & 0.1606$^{a}$
 & 0.5467
 \\
\midrule
LLaMA-3-8B
 & 0.6602
 & 0.1569$^{a}$
 & 0.4504$^{a}$
 & 0.1058$^{a}$
 & 0.3235$^{a}$
 & 0.4011
 \\
LLaMA-3.1-8B
 & 0.6949
 & 0.1619$^{a}$
 & 0.5167$^{a}$
 & 0.1340$^{a}$
 & 0.3261$^{a}$
 & 0.4102
 \\
LLaMA-3.3-70B
 & 0.7286
 & 0.0741$^{a}$
 & 0.2973$^{a}$
 & 0.0978$^{a}$
 & 0.1052$^{a}$
 & 0.5850
 \\
\midrule
Flan-T5-Large (0.8B)
 & 0.6601
 & 0.6609
 & 0.6845$^{a}$
 & 0.6553
 & 0.6807
 & -0.0102
 \\
Flan-T5-XL (3B)
 & 0.6949
 & \textbf{0.6910}
 & \textbf{0.6994}
 & \textbf{0.6934}
 & \textbf{0.7142}$^{a}$
 & -0.0046
 \\
\bottomrule
\end{tabular}
\end{table}
\subsubsection{\textbf{Domain Robustness}}
Figure ~\ref{fig:rq6_full}(a) reveals substantial consistency in attack success rates between General and Domain-specific datasets across all three ranking schemas. Pairwise and setwise configurations demonstrate the highest susceptibility to attacks, whereas listwise ranking exhibits lower vulnerability. 

\textbf{These findings establish that vulnerability transcends domain boundaries}. The close alignment between general and domain-specific performance bars demonstrates that preference attacks retain their potency when transferred to specialized benchmarks. The scatter plot in Figure~\ref{fig:rq6_full}(b) provides compelling quantitative evidence: plotting each model's TREC-DL ASR (x-axis) against its BEIR ASR (y-axis) yields a Pearson correlation coefficient of $r = 0.969$ ($p = 0.0003 < 0.05$), indicating an exceptionally strong and statistically significant linear relationship between attack effectiveness on general versus specialized datasets. This robust correlation confirms that \textbf{vulnerability appears to be primarily an intrinsic model property}, with dataset-specific effects playing a secondary modulatory role.
The coefficient of determination ($R^2 = 0.938$) further substantiates this conclusion: approximately 93.8\% of the variance in domain-specific attack success rates can be predicted from general TREC-DL performance alone. 
While the correlation is remarkably strong, a subtle asymmetry emerges: all LLMs fall marginally below the $y = x$ identity line, indicating that attacks achieve slightly higher success rates on TREC-DL than on BEIR datasets. The heatmap in Figure~\ref{fig:rq6_full}(c) elucidates this pattern, showing that TREC-DL datasets generally exhibit elevated ASR values compared to BEIR datasets.

More granular analysis of the heatmap (Figure~\ref{fig:rq6_full}(c)) reveals heterogeneous vulnerability across benchmarks, driven primarily by \textit{textual length} and \textit{relevance levels}. 
DBpedia emerges as the most susceptible dataset, characterized by the shortest average document length in our study ($49.7$ words). Compared to the MS MARCO passage average of $56.0$ words, DBpedia's constrained context correlates with higher ASRs across all models. A plausible underlying mechanism is the textual dilution effect: in shorter documents, the injected instruction occupies a larger relative proportion of the input, potentially increasing its attention salience. This observation is further supported by the comparison between TREC-COVID ($160.8$ words) and Touché-2020 ($292.4$ words). Despite sharing the same multi-level relevance levels, the longer documents in Touché-2020 consistently exhibit lower ASRs.

Beyond textual volume, the granularity of relevance levels appears to be a critical modulator of robustness. A key discrepancy arises when comparing SciFact ($213.6$ words) and Touché-2020: despite featuring shorter documents, SciFact maintains lower ASR values, indicating higher relative resistance. This anomaly is likely rooted in the relevance level of the underlying datasets. SciFact employs binary relevance assessments, whereas datasets like Touché-2020 or TREC-DL utilize three- or four-level graded assessments. Since the ASR evaluation protocol constructs candidate sets by sampling from distinct relevance tiers, SciFact's binary structure results in significantly lower relevance heterogeneity.

\begin{figure}
	\centering
	\includegraphics[width=1\columnwidth]{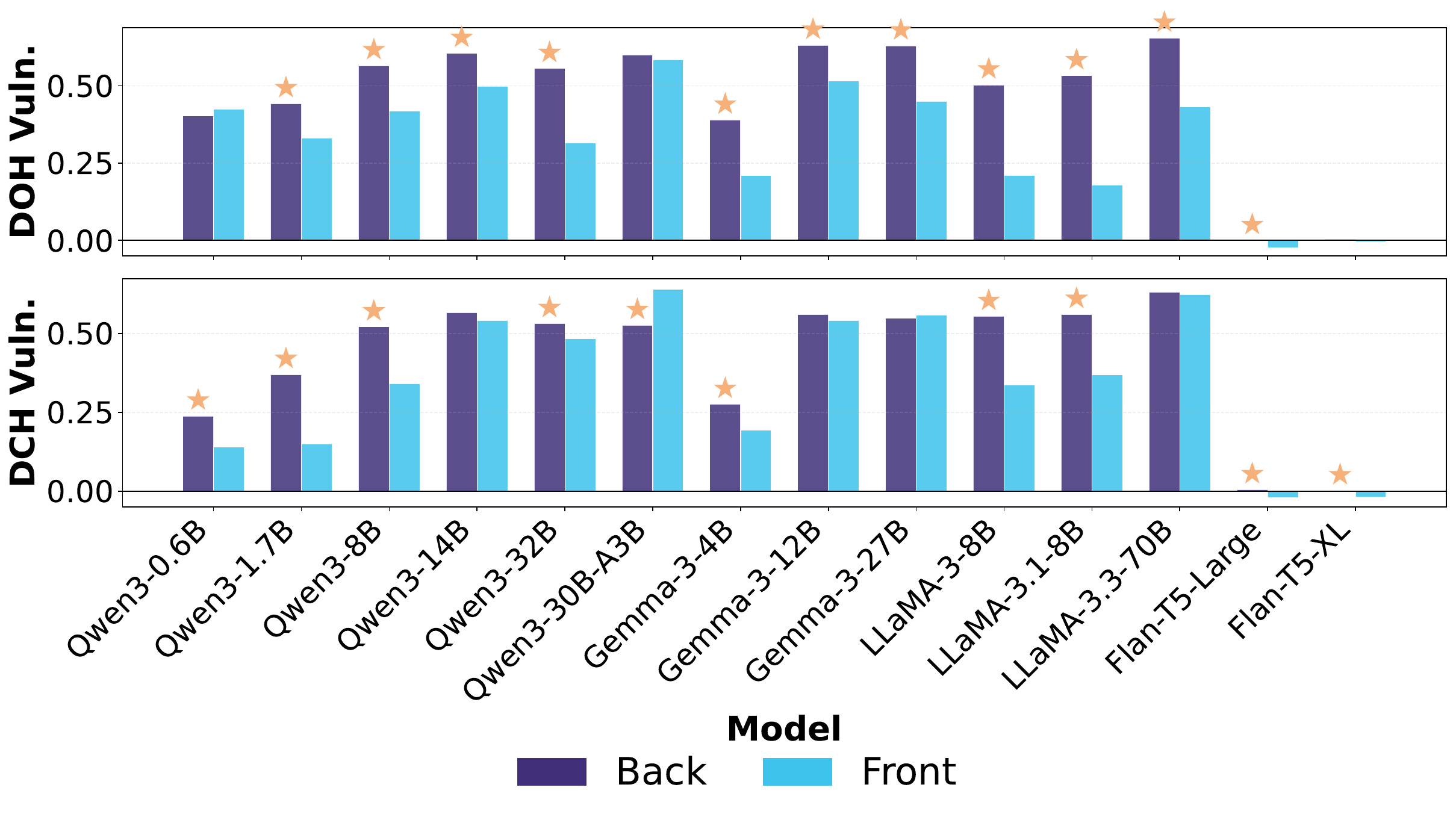}
	\caption{\textbf{Comparative analysis of model vulnerability across injection positions.} $\star$: statistically significant position effect.}
	\label{fig:rq7_position}
\end{figure}

\subsection{Task2: Retrieval Effectiveness Assessment}

Table~\ref{tab:compact_eval_half} presents a comprehensive analysis of ranking performance degradation under DCH and DOH attacks across all evaluated LLM backbones and injection positions using setwise ranking. 

\subsubsection{\textbf{Reproducibility}}
While Qian et al. do not report setup details, our setwise measurements are broadly comparable to their published results (Appendix A2 of the original study): for LLaMA-3-8B, we obtain a baseline nDCG@10 of 0.6602 (vs. 0.6930) and, when attacked, a drop to 0.1569 in our DOH-back condition (vs. 0.1050).

\subsubsection{\textbf{Vulnerability Scale}}
Results indicate that high ASR translates into substantial ranking loss. We observe a dramatic baseline performance drop under every attack setting. Averaged across attack settings, mean nDCG@10 drops from 0.6584 to 0.2947 (55.2\% relative decline).
Unlike ASR, nDCG@10 serves as a composite metric for both ranking ability and adversarial robustness. Notably, it reveals that \textbf{larger and more capable LLMs suffer from more severe attack-induced degradation}.
Note, we excluded Gemma-3-1B due to its failure to produce valid output, causing it to largely under-perform BM25 even when no attacks are present. For the remaining models, the performance gaps between baseline and various attacks are statistically significant in almost all cases, and the attacks are so powerful that they result in ranker performance below that of the first-stage BM25.

\begin{figure}
	\centering
	\includegraphics[width=1\columnwidth]{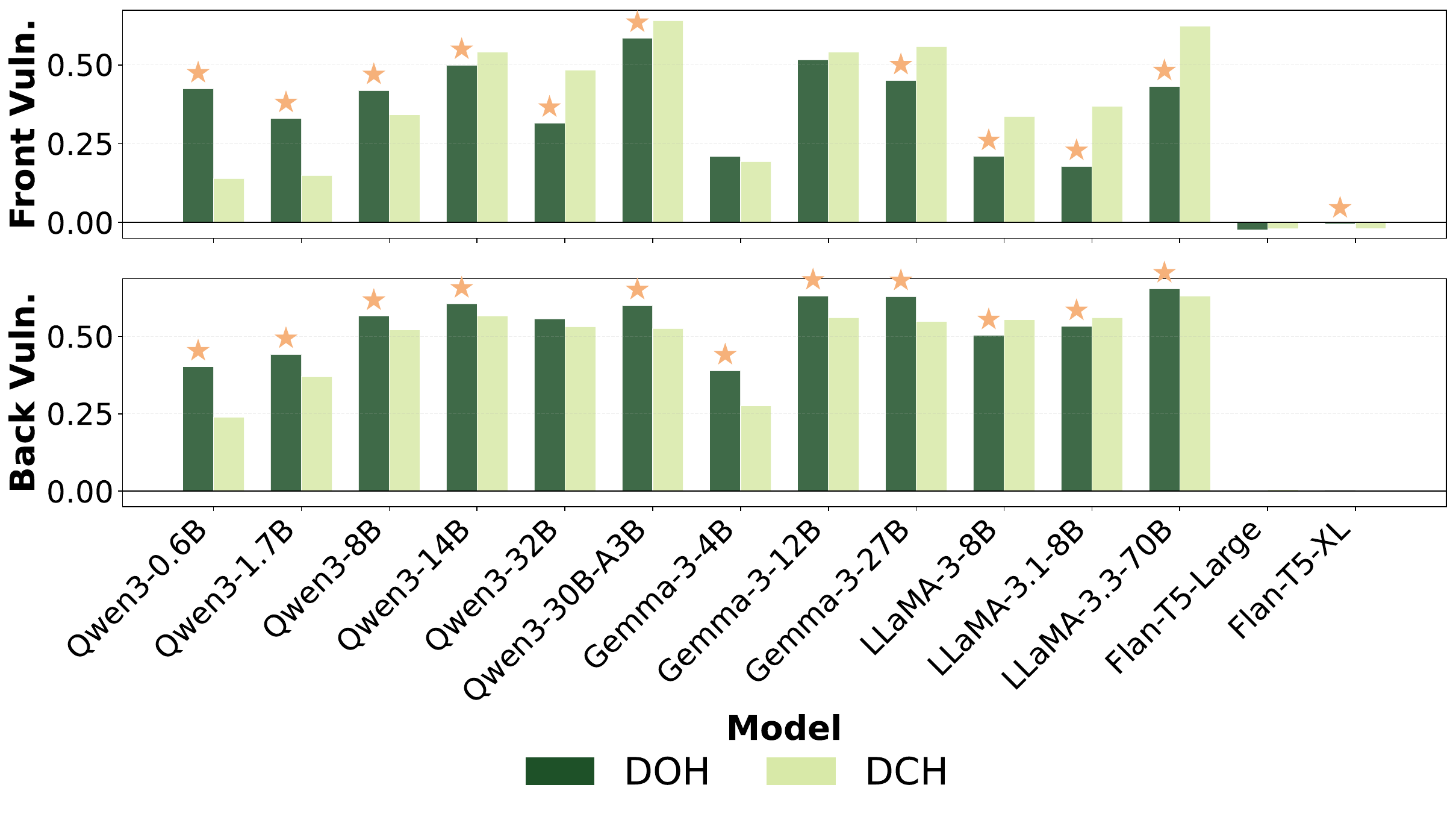}
	\caption{Comparative analysis of model vulnerability across attack methods. $\star$: statistically significant method effect.}
	\label{fig:rq7_method}
\end{figure}

\subsubsection{\textbf{Position Sensitivity}}
Figure~\ref{fig:rq7_position} shows that position effects are even more pronounced when measured in terms of ranking effectiveness losses than by raw ASR under both attack settings. In particular, for 14 of the 15 evaluated LLM backbones, back-placed injections lead to significantly larger performance degradation than front-placed ones, with a mean nDCG@10 gap of 0.1106. Furthermore, the prevalence of statistically significant back--front differences across both attack settings (paired $t$-test, $p<0.05$) indicates that for setwise ranking where the injection is placed is a decisive determinant of practical harm: back placement is generally more disruptive to the final ranking quality.

\subsubsection{\textbf{Architectural Divergence}}
We observe strong LLM architecture-dependent differences in downstream robustness. The Flan-T5 family exhibits exceptional resilience, consistently achieving the lowest degradation across the board. Intriguingly, Flan-T5 models show negative degradation (i.e., slightly improved $nDCG@10$ under attack compared to the baseline). While Flan-T5 models still exhibit non-zero ASR, the successful ``hijacks'' are apparently insufficient to degrade the overall ranking quality. The structural robustness of encoder-decoder architectures makes them prime candidates for building more secure ranking systems. Conversely, the MoE-based Qwen3-30B-A3B suffered the highest overall $nDCG@10$ degradation, consistent with ASR results.

\subsubsection{\textbf{Attack Methods Sensitivity}}
We further examine whether DOH and DCH differ in their end-to-end impact. Figure~\ref{fig:rq7_method} compares nDCG@10 degradation under the two attack mechanisms. Although the mean difference between DOH and DCH is marginal (0.0090), the paired comparisons are frequently statistically significant in the setwise ranking setting (paired $t$-test, $p<0.05$). This finding suggests that the two methods exert distinct, systematic influences on downstream ranking quality, implying that even a subtle utility gap can reflect fundamentally different failure modes.

\begin{figure}
	\centering
	\includegraphics[width=0.85\columnwidth]{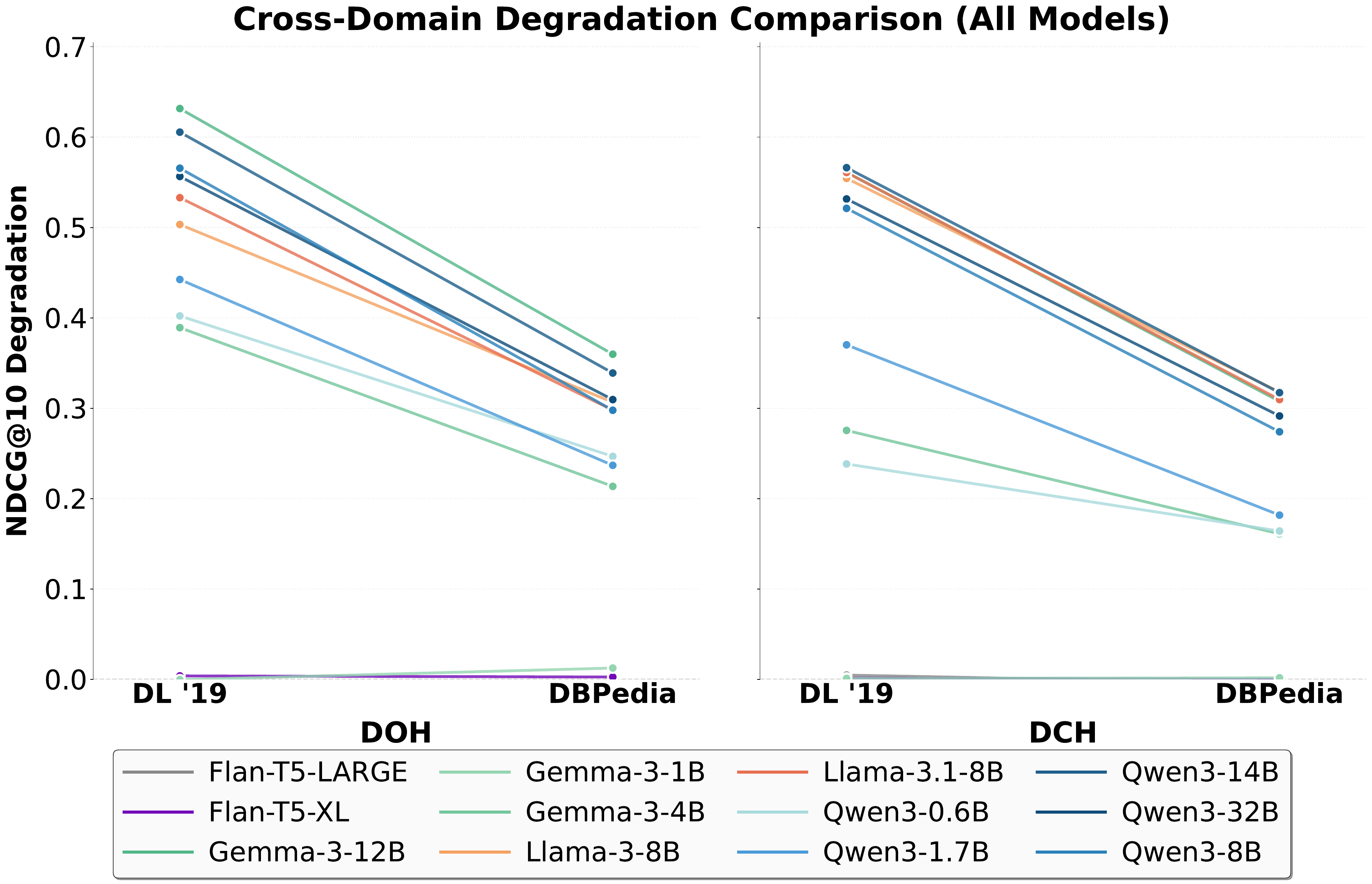}
	\caption{NDCG@10 degradation comparison between TREC-DL 2019 and DBPedia: the higher the degradation, the worse the NDCG due to the attack.}
	\label{fig:rq7_degradation_slope}
\end{figure}

\subsubsection{\textbf{Domain Robustness}}
Figure \ref{fig:rq7_degradation_slope} reveals a notable discrepancy between ASR and nDCG@10: although DBpedia exhibits consistently higher ASR (Figure~\ref{fig:rq6_full}(c)), TREC-DL 2019 experiences substantially more severe nDCG@10 degradation. This phenomenon can be attributed to vulnerability scaling with model capacity—higher-performing models demonstrate steeper performance declines under attack, with comprehensive results available in our repository. Notably, Flan-T5 models consistently exhibit remarkable cross-domain robustness with negligible degradation.


	\section{Conclusion}
	This paper comprehensively revisits and extends \citet{qian2025ranking}'s study on attacking LLM rankers through a focused reproducibility study and a set of systematic analyses that connect intrinsic vulnerability to end-to-end retrieval impact. By decoupling our analysis into two complementary tasks— preference vulnerability assessment (Task 1) and ranking vulnerability assessment (Task 2)—we bridge the gap between model susceptibility and its operational consequences for ranking. Our results confirm the core findings of the original work across differing environments, supporting strong overall reproducibility.

Synthesizing evidence from both tasks, we delineate four boundary conditions of decision-hijacking vulnerabilities (see Table~\ref{tab:research_dimensions} for a succinct overview):

\begin{enumerate}[leftmargin=14pt]
	\item We confirm the original scaling claim: larger, more capable LLMs are generally more vulnerable.
	
	\item We show that attack placement is important, unlike what reported by Qian et al.: in setwise ranking, back-of-passage injections are frequently more damaging than front-of-passage ones, with differences that are frequently statistically significant.
	
	\item We uncover a marked architectural divergence: encoder-decoder models (Flan-T5) are substantially more robust than decoder-only LLMs, exhibiting much lower ASR and negligible degradation.
	
	\item We find that vulnerability largely generalizes across domains, indicating that susceptibility is primarily an intrinsic model property, while also being modulated by dataset factors such as text length and relevance-level structure.
\end{enumerate}

 
 These findings characterize the boundary conditions of jailbreak prompt vulnerabilities, revealing critical insights and offering actionable guidance for designing robust neural retrieval systems.  Reflecting more broadly, we note that the attacks considered by Qian et al. and also in this paper involve having knowledge of which are the non-relevant candidates retrieved by the first stage of the ranking pipeline and having the ability to inject the attack on all these documents: conditions that are often unlikely to occur in practice. We believe that future work should be directed in designing more realistic attack conditions which better align with what might be achievable by malicious agents wanting to deteriorate the ranker performance (like in the unrealistic settings considered here) or performing SEO activities to boost the rank of target pages (that might or not be relevant to the user query). 
	
	\bibliography{custom}
	
\end{document}